\documentclass[12pt,a4paper]{article}%

\usepackage{amsmath,amssymb,amsfonts}
\usepackage{calc}
\usepackage{caption2}
\usepackage{hyperref}
\usepackage[pdftex]{color,graphicx}
\usepackage{cite}
\usepackage{multirow}

\numberwithin{equation}{section} 
\setcaptionmargin{1cm}

\long\def\symbolfootnote[#1]#2{\begingroup
\def\thefootnote{\fnsymbol{footnote}}\footnote[#1]{#2}\endgroup}

\graphicspath{{fig/}}

\begin{document}
\begin{titlepage}
\vskip 1.0cm
\begin{center}
{\Large \bf Detecting the Higgs boson(s) in $\lambda$SUSY  }
\vskip 1.0cm {\large  Enrico Bertuzzo$^{\,a,b}$, Marco Farina$^{\,a}$} \\[1cm]
{\it $^{a}$ Scuola Normale Superiore and INFN, Piazza dei Cavalieri 7, 56126 Pisa, Italy} \\[3mm]
{\it $^{b}$ Institut de Physique Th\'eorique\!
\symbolfootnote[4]{Laboratoire de la Direction des Sciences de la
Mati\`ere du Commissariat \`a l'Energie Atomique et Unit\'e de
Recherche associ\'ee au CNRS (URA 2306).}, CEA-Saclay,\\
  F-91191 Gif-sur-Yvette Cedex, France.}
\vskip 1.0cm
\today
\end{center}

\begin{abstract}
We reconsider the Higgs bosons discovery potential in the $\lambda$SUSY framework, in which the masses of the scalar
particles are increased already at tree level via a largish supersymmetric coupling between the usual Higgs doublets and a
Singlet. We analyze in particular the interplay between the discovery potential of the lightest and of the next-to-lightest
scalar, finding that the decay modes of the latter should be more easily detected at the LHC.
\end{abstract}
\end{titlepage}

\section{Introduction}

The quest for the nature of the Electroweak Symmetry Breaking (EWSB) mechanism is reaching its most important phase:
with an integrated luminosoty of $2.3\;{\rm fb^{-1}}$, the combined CMS-Atlas data exclude at $95\%$ C.L. a Standard Higgs boson mass in the $(141\div 476)\;{\rm GeV}$ range \cite{CMS-Atlascomb}.\\
Of course, since in many extensions of the Standard Model (SM) the Higgs boson is far from being standard
(both in its production and decay modes), the presence of such a scalar triggering EWSB is not yet excluded.
\par
Supersymmetry is surely one of the best motivated extension of the SM, since it stabilizes the Electroweak scale against radiative
corrections. Despite this, in the Minimal Supersymmetric Standard Model (MSSM),
the tree level Higgs boson mass is bounded from above
at most by the Z boson mass ($m_h \leq m_Z |\cos 2\beta|$), so that one must rely on considerable radiative corrections in order to satisfy the LEP bound. At the same time, since in
wide regions of the parameter space the MSSM Higgs boson has standard couplings, it is now highly constrained by the LHC. This
may be an indication that the theory must be augmented with an additional Singlet field, whose coupling $\lambda$ with the two
standard Higgs bosons allows to increase the previous upper bound to
$m_h^2 \leq m_Z^2 \cos^2 2\beta + \lambda^2 v^2 \sin^2 2\beta$.
\par
This case was studied extensively in the literature, known as Next to Minimal Supersymmetric Standard Model
(NMSSM, for comprehensive reviews, see \cite{Ellwanger:2009dp, Maniatis:2009re}), with a superpotential given by $W= \lambda S H_1 H_2 + (k/3) S^3$.
Not only the Singlet field allows one to increase the tree level upper bound on the Higgs boson mass but also the scalars production and decay rates are changed, since in general they will now have a Singlet component that does not
interact with the SM matter or gauge fields.
\par
A crucial point for the phenomenology of the model is the value of the trilinear coupling $\lambda$.\\
The most studied situation is the one in which perturbativity is retained up to the GUT scale, so that at low energy
an upper bound $\lambda \lesssim 0.7$ holds. The mass of the lightest scalar can be below the LEP bound, but we no longer need it to be above $114\;{\rm GeV}$ since its couplings are no longer standard,
both in production and in decay. What is interesting is that there can be a sort of inversion of roles between the lightest and the next-to-lightest scalar,
with the latter that can be more similar (in a sense to be made more precise later) to the SM Higgs boson. This ``No-Lose'' theorem in which at least
one of the NMSSM scalars should be discovered at the LHC relies on the assumption that Higgs-to-Higgs or Higgs-to-SUSY decays are
kinematically not allowed \cite{Hugonie:2001ib, Ellwanger:2001iw, Ellwanger:2003jt, Miller:2004uh, Ellwanger:2005uu, Moretti:2006hq, Forshaw:2007ra,
Belyaev:2008gj, Belyaev:2010ka, Almarashi:2010jm, Almarashi:2011hj}, while important violations are possible once one or both the previous
assumptions are relaxed \cite{Ellwanger:1999ji, Barbieri:2007tu, Djouadi:2008uw, Mahmoudi:2010xp}.
\par
Dropping the requirement of perturbativity up to the GUT scale, $\lambda$ can take larger values at the EW scale
(\emph{i.e} requiring perturbativity up to $10\;{\rm TeV}$ increases the low energy upper bound to $\lambda \lesssim 2$). Since in this case the tree level mass of the scalars can take values up to $(200\div 250)\;{\rm GeV}$,
naturalness is improved. This situation has been recently studied in \cite{Franceschini:2010qz, Cao:2008un, Bertuzzo:2011ij}, where it has been shown that the behavior of the lightest scalar in this
``$\lambda$SUSY'' framework\footnote{We call this framework $\lambda$SUSY to stress the importance of the higher value of
the $\lambda$ coupling.} can be quite similar to the one of the usual NMSSM\footnote{Although there are also regions in which it is simply a heavier standard Higgs boson
(resembling what happens in other realizations of $\lambda$SUSY, see \cite{Barbieri:2006bg, Cavicchia:2007dp}) and is thus now excluded.}.
\par
It is then interesting to see whether the ``No-lose'' theorem, which seems to be valid under certain assumptions for the NMSSM with small $\lambda$, can
be somehow extended to the $\lambda$SUSY case. This seems to be possible at least in ample regions of the parameter space, as we
will see in more detail in the next Section.

\section{Notation and plots}
\begin{center}
\begin{table}[tb]
\centering
 \begin{tabular}{c|ccc|cccc|c|c}
\multirow{2}{*}{$k$} & \multicolumn{3}{|c|}{Masses  (GeV) } & \multicolumn{4}{|c|}{BRs} &
\multirow{2}{*}{$\frac{\sigma(pp\rightarrow s)}{\sigma(pp\rightarrow h_{SM})}$} &
\multirow{2}{*}{$\xi$}\\
 & $m_{A_1}$ & $m_{\chi_1}$ & $m_s$ & $A_1A_1$ &$Z A_1$ & $\chi_1\chi_1$ &$WW$ &\\
\cline{1-10}
\multirow{6}{*}{-0.2} &\multirow{2}{*}{103} & \multirow{2}{*}{130} & 252 & 0.54 & 0.01 & 0 & 0.31  & 0.38 &  0.17\\
 & & & 284 & 0.032 & 0.324 & 0.043 & 0.41 & 1.06 & 0.62\\
\cline{2-10}
 &\multirow{2}{*}{95} & \multirow{2}{*}{77} & 163 & 0 & 0 & 0.8 & 0.06    & 0.56 & 0.04\\
 & & & 204 & 0.4 & 0 & 0.143 & 0.33  & 0.82 & 0.37  \\
\cline{2-10}
&\multirow{2}{*}{108} & \multirow{2}{*}{96} & 173 & 0 & 0 & 0 & 0.79   & 0.69 & 0.57\\
 & & & 243 & 0.412 & 0 & 0.086 & 0.35 & 0.70 & 0.35\\
\cline{1-10}
\multirow{6}{*}{-0.6} &\multirow{2}{*}{166} & \multirow{2}{*}{78} & 160 & 0 & 0 & 0.72 & 0  & 0.38 & $10^{-4}$\\
 & & & 194 & 0 & 0 & 0.189 & 0.61 & 1 & 0.8\\
\cline{2-10}
 &\multirow{2}{*}{195} & \multirow{2}{*}{120} & 232 & 0 & 0 & 0 & 0.69 & 0.04 & 0.04\\
 & & & 248 & 0 & 0 & 0.001 & 0.70 & 1.4 &  1.3  \\
\cline{2-10}
&\multirow{2}{*}{168} & \multirow{2}{*}{133} & 218 & 0 & 0 & 0 & 0.71 & 0.52 & 0.5\\
 & & &  318 & 0 & 0.21 & 0.145 & 0.44 & 0.92 & 0.6\\
\cline{1-10}
\end{tabular}
\caption{\label{Tab:xi} \small Masses and Branching Fractions for some relevant points in parameter space. The first row in each group always refers to $s_1$, the second one to $s_2$.
The mass of $s_3$ is always larger than about $500\;{\rm GeV}$, while the second pseudoscalar has mass large enough to never be relevant in the decay channels of the particles
of interest. The parameters are chosen as follows: $\lambda=2$, $\tan\beta=1.5$; for $k=-0.2$ the points refer respectively to $(\mu\; ({\rm GeV}), m_H\; ({\rm GeV}))= (180, 340), \;(105, 180), \;
(130, 200)$, while for $k=-0.6$ they refer to $(\mu\; ({\rm GeV}), m_H\; ({\rm GeV}))= (105, 180), \; (160, 280), \; (180, 370)$.
 In the last two columns we present the ratio between the production cross section via gluon-gluon fusion for the relevant scalar and a Standard Model Higgs boson \cite{Bertuzzo:2011ij, Franceschini:2010qz} and $\xi$
(Eq. \ref{eq:xi}).}
\end{table}
\end{center}
Let us now briefly explain our notation (for a detailed discussion about the scalar potential and the parameter space
that we are considering we refer the reader to \cite{Cao:2008un, Franceschini:2010qz, Kanehata:2011ei, Bertuzzo:2011ij}). We define our model through
\begin{eqnarray}
 W &=& \lambda S H_1 H_2+ \frac{k}{3} S^3 \;,\nonumber\\
 V_{SSB} &=& m_1^2|H_1|^2+ m_2^2 |H_2|^2 + m_S^2 |S|^2 -\left(\lambda A S H_1 H_2 + \frac{k G}{3} S^3+h.c.\right)\;,
\end{eqnarray}
with scalar particles defined by
\begin{equation}
 H_{1,2} = v_{1,2} + \frac{h_{1,2}+i a_{1,2}}{\sqrt{2}}, ~~~S= v_s + \frac{S_1+i S_2}{\sqrt{2}}.
\end{equation}
The model has $7$ parameters ($\lambda$, $k$, $m_1^2$, $m_2^2$, $m_S^2$, $A$, $G$), that can be reduced to $6$ using the minimum conditions and
considering that one combination is fixed by the vev $v=\sqrt{v_1^2+v_2^2} \simeq 174\;{\rm GeV}$. We are left with
\begin{equation}
 \lambda, ~~k, ~~\tan\beta, ~~ \mu, ~~ m_H, ~~~A/G
\end{equation}
 where $\mu=\lambda v_s$ is the higgsino mass parameter and $m_H$ the charged Higgs boson mass. In what follows we will choose for simplicity $A/G=1$,
but we have checked that changing this value does not affect in a consistent way our conclusions.
\par
In the neutralino sector there is also a dependence on the Majorana gaugino masses $M_{1,2}$. For concreteness, in the following we will set $M_2 = 2\;{\rm TeV}$ and $M_1 = 200\;{\rm GeV}$ having in mind a possible
bino-higgsino well-tempered DM candidate \cite{ArkaniHamed:2006mb}\footnote{This can be justified since the singlino component of the LSP is always below $10\%$ for our choice
of parameters.}.
\par
We choose to present our results in the $(\mu, m_H)$ plane to deal with physical parameters. The allowed regions in parameter space are found considering properly
the minimum conditions and the requirement of CP conservation in the scalar sector \cite{Franceschini:2010qz, Bertuzzo:2011ij}.
\par
In what follows we will always take $\lambda=2$, $|k| \lesssim 0.7$ and $\tan\beta=1.5$. Such a large value for $\lambda$ is the characteristic feature of our
model, since it allows to increase the Higgs boson mass up to $(200\div250)\;{\rm GeV}$. We are then requiring semiperturbativity up to $10\;{\rm TeV}$ or so, and
this motivates the upper bound on $k$ (for $k\simeq 0.7$ at the EW scale this coupling becomes semiperturbative  at $10\;{\rm TeV}$). The rather low value for $\tan\beta$ is instead due
to requirement of consistency with the ElectroWeak Precision Tests. In what follows, we will
choose two representative values for $k$, $k=-0.2$ and $k=-0.6$: for intermediate values the overall conclusion is similar, although a few details may change.
\par
The physical scalar particles will be called $s_{1,2,3}$ in the CP-even sector and $A_{1,2}$ in the CP-odd sector, all
named in increasing order of masses. For our discussion, $s_3$ and $A_2$ are not significant, since the mass of the heaviest scalar is always
above $500\;{\rm GeV}$ while the mass of $A_2$ is large enough to be irrelevant in the decays of the two lightest scalars.
\par
The crucial quantity we want to analyze is
\begin{equation}\label{eq:xi}
 \xi_i = \frac{\sigma(pp\rightarrow s_i) {\rm BR}(s_i \rightarrow VV)}{\sigma(pp\rightarrow h_{SM}) {\rm BR}(h_{SM} \rightarrow VV)}
\end{equation}
with $h_{SM}$ the usual Standard Model Higgs particle and $V$ any of the vector bosons, since these channels are the most sensitive for the range of masses of
interest.
\par
\begin{center}
 \begin{figure}[tb]
  \begin{tabular}{cc}
   \includegraphics[width=0.45\textwidth]{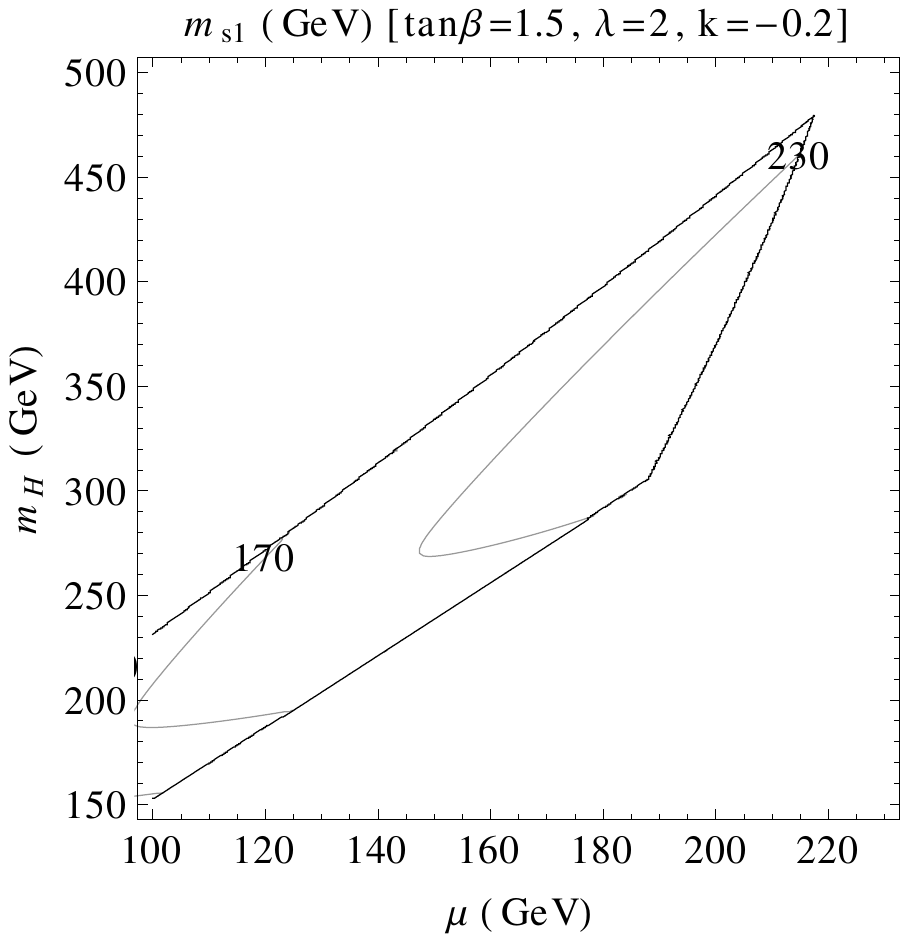} & \includegraphics[width=0.45\textwidth]{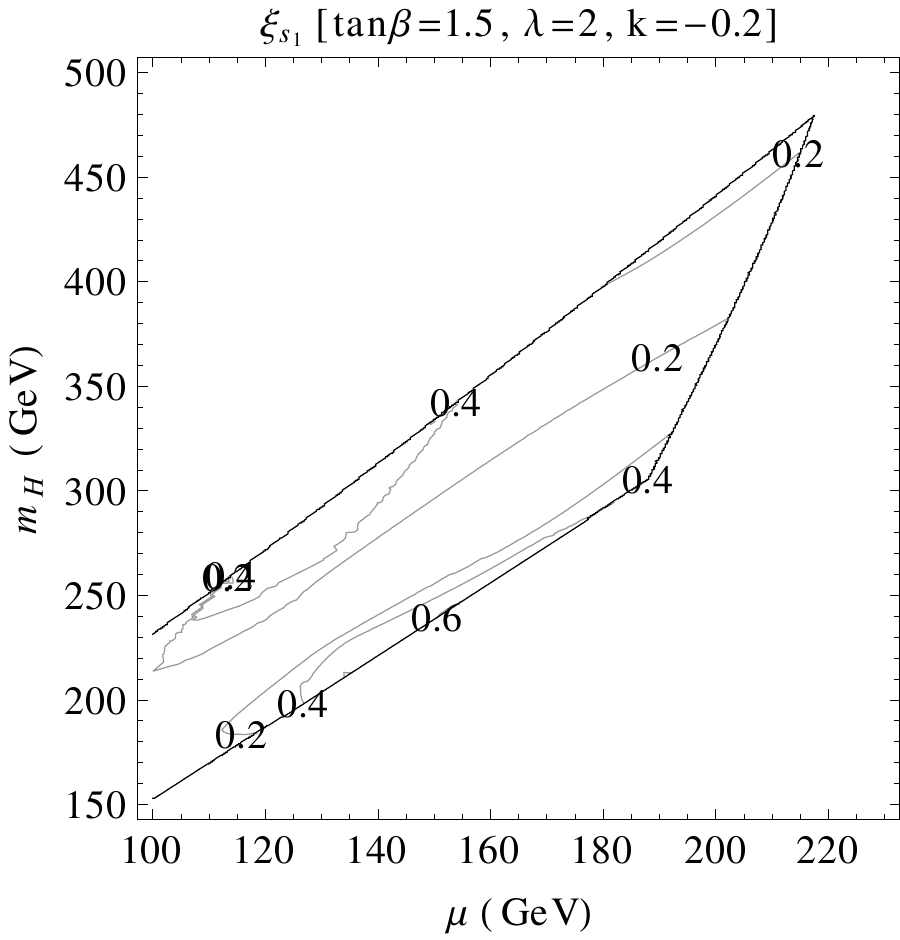} \\
   \includegraphics[width=0.45\textwidth]{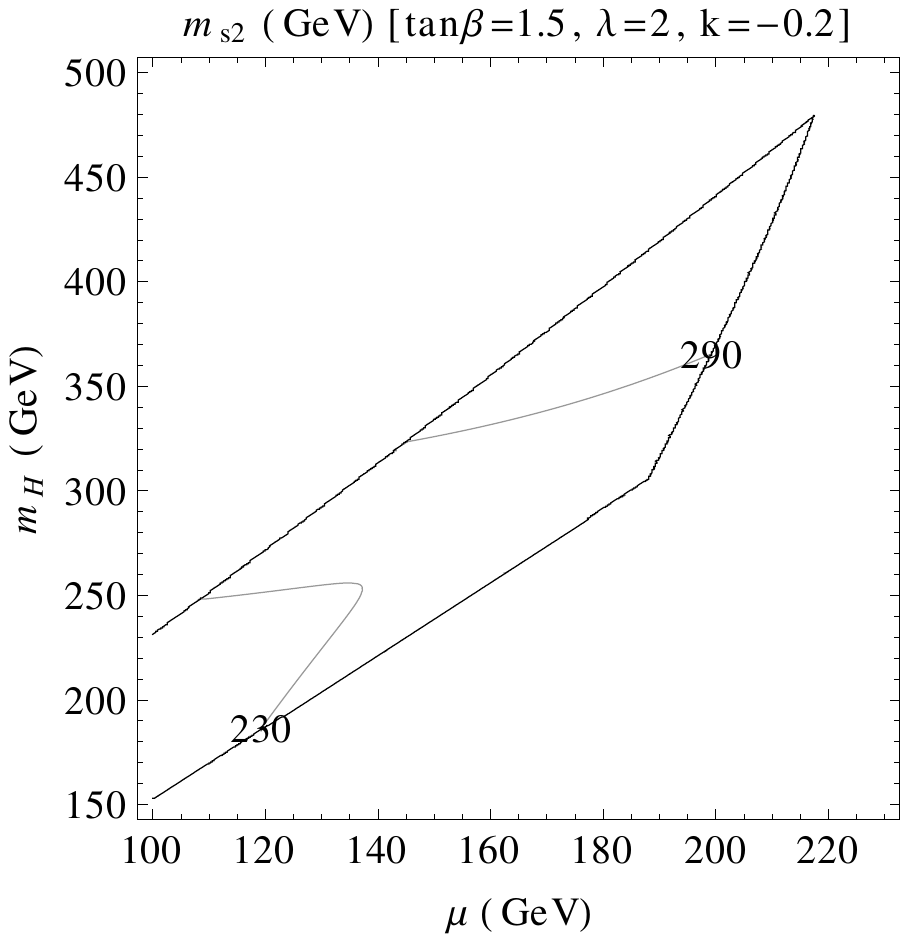} & \includegraphics[width=0.45\textwidth]{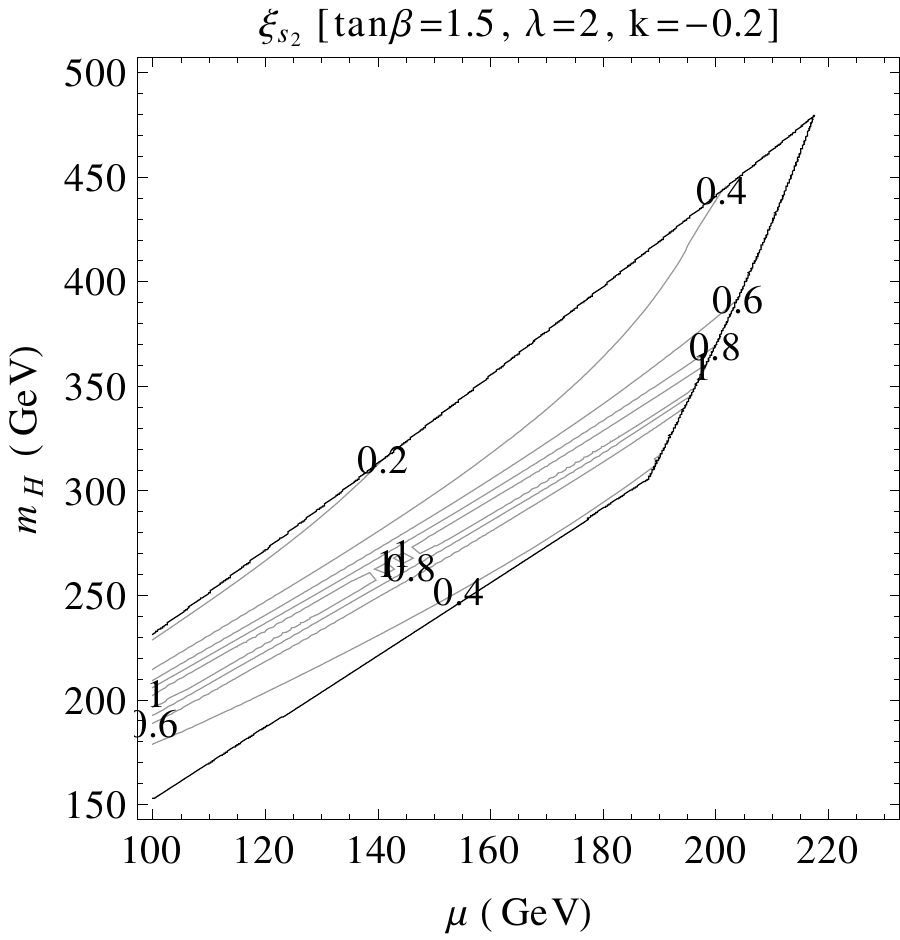}
  \end{tabular}
\caption{\label{fig:k02} \small Isolines of the mass (left panels) and of $\xi$ (right panels) for the two lightest scalars in the $k=-0.2$ case. The other parameters are fixed as: $\lambda=2$, $\tan\beta=1.5$,
$M_1=200\;{\rm GeV}$, $M_2=2\;{\rm TeV}$.}
 \end{figure}
\end{center}
\begin{center}
 \begin{figure}[tb]
  \begin{tabular}{cc}
   \includegraphics[width=0.45\textwidth]{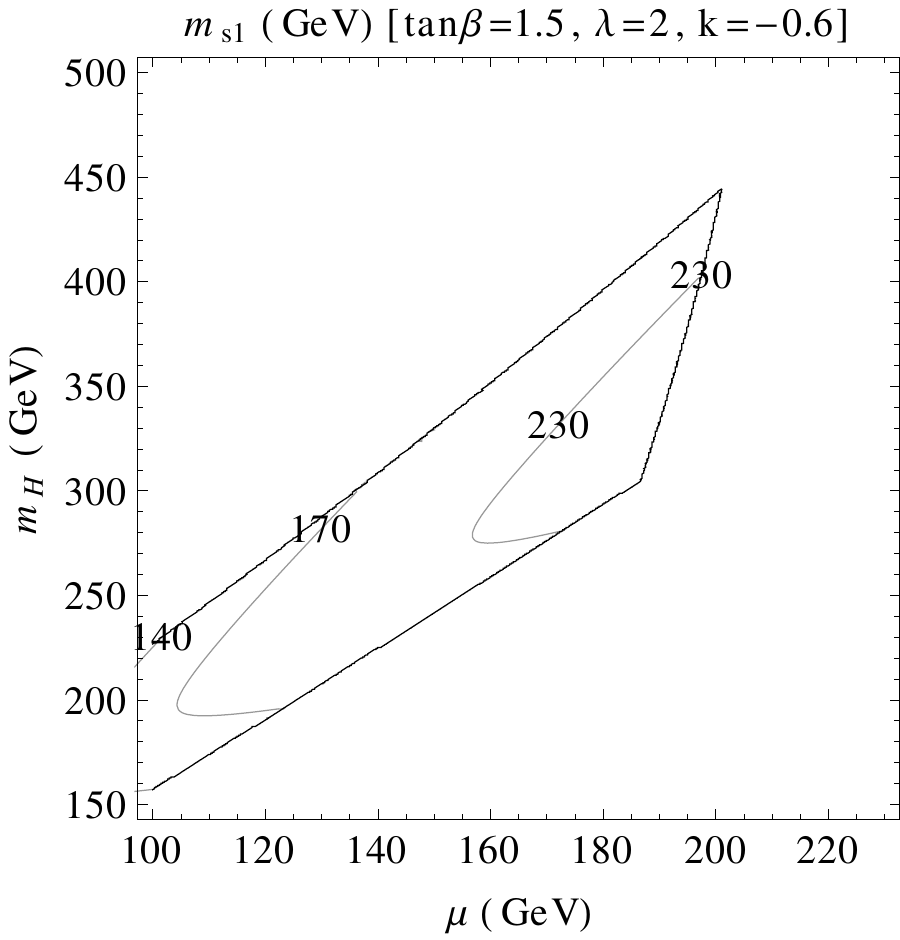} & \includegraphics[width=0.45\textwidth]{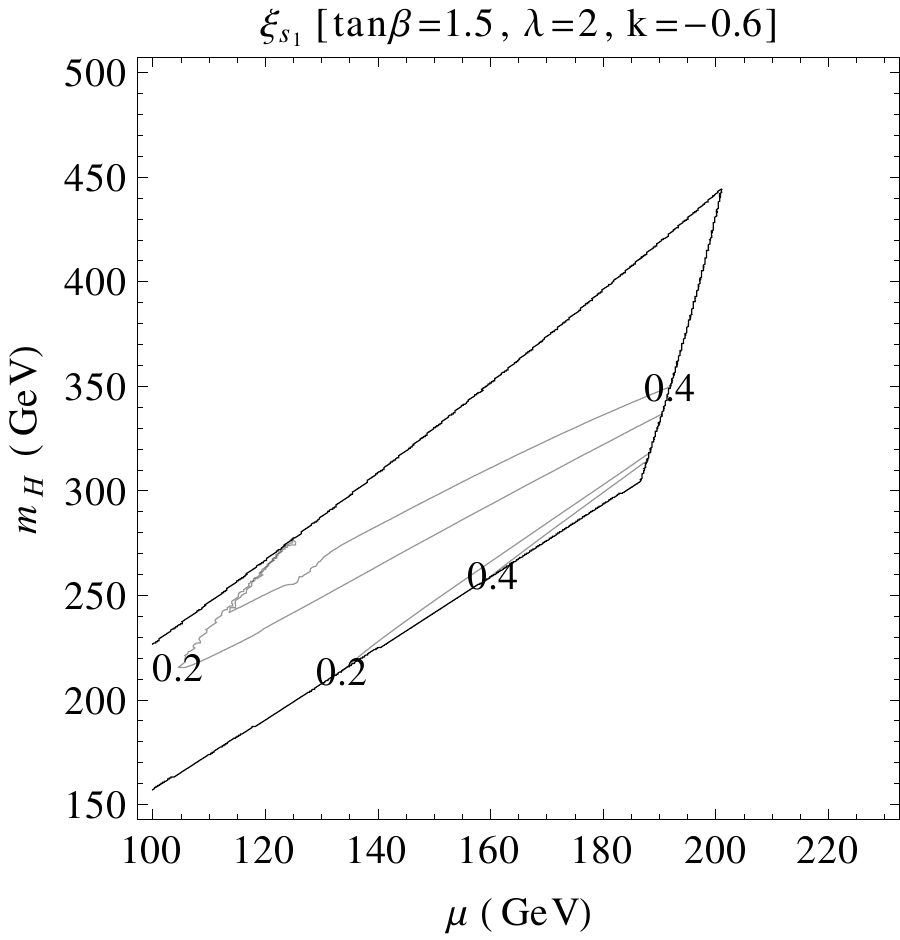} \\
   \includegraphics[width=0.45\textwidth]{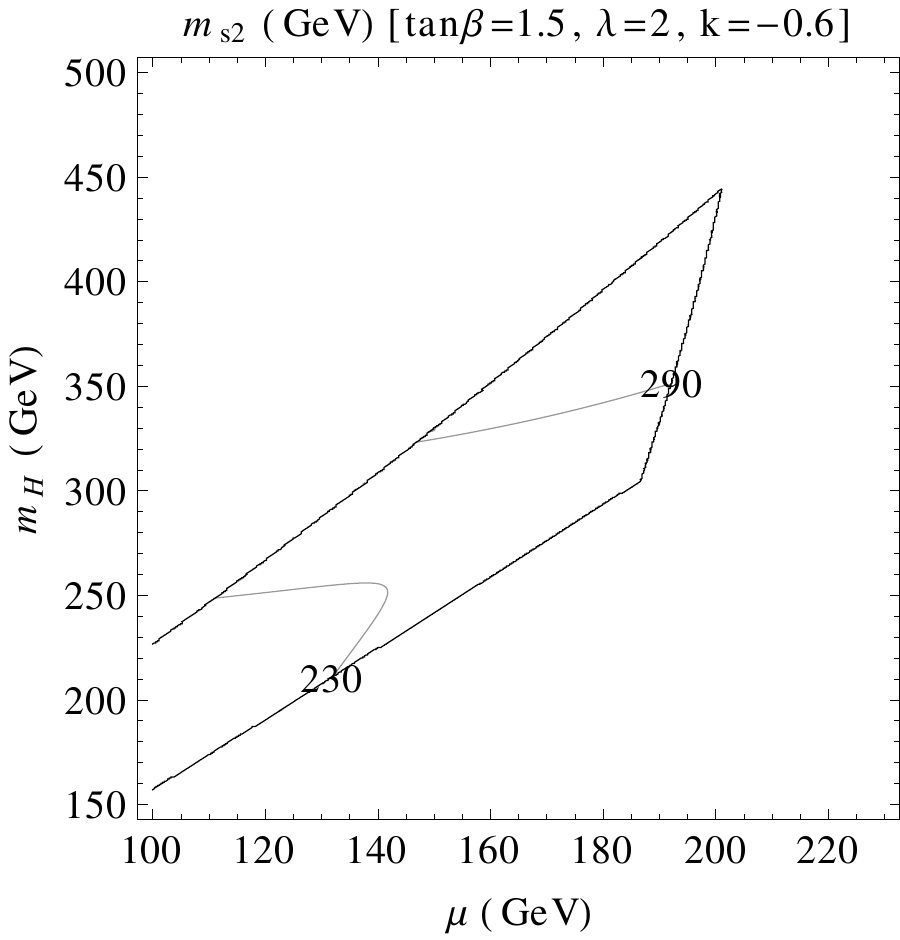} & \includegraphics[width=0.45\textwidth]{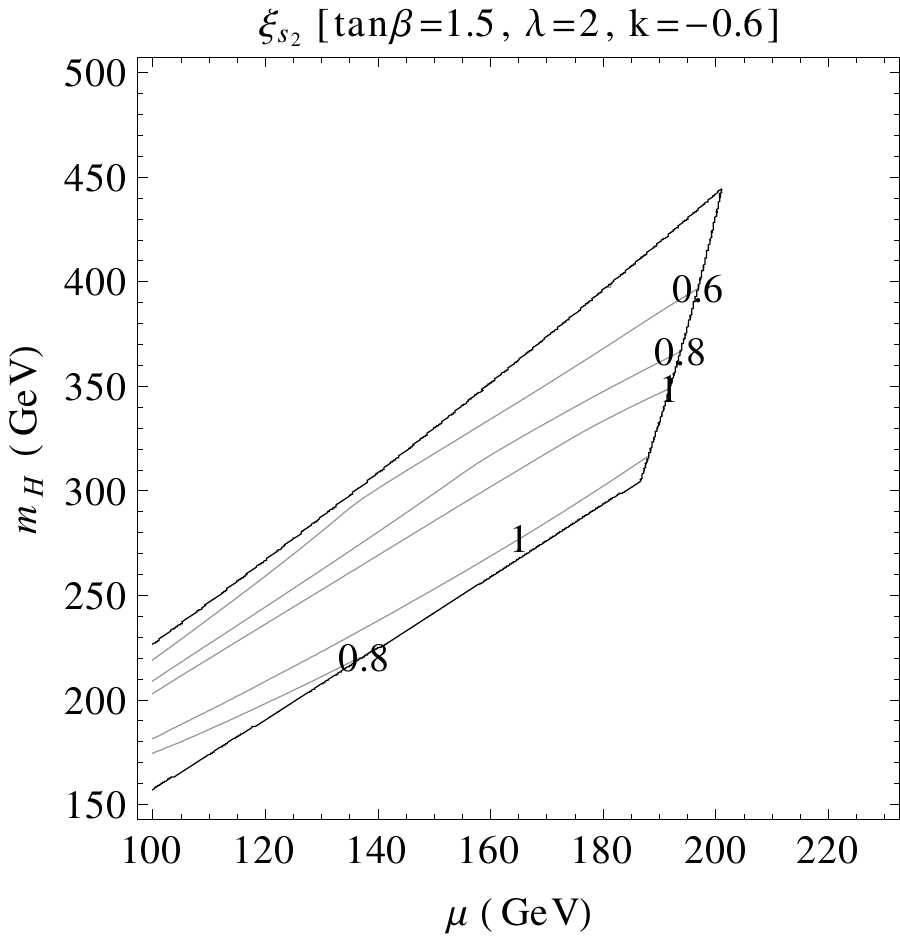}
  \end{tabular}
\caption{\label{fig:k06}\small Isolines of the mass (left panels) and of $\xi$ (right panels) for the two lightest scalars in the $k=-0.6$ case. The other parameters are fixed as: $\lambda=2$, $\tan\beta=1.5$,
$M_1=200\;{\rm GeV}$, $M_2=2\;{\rm TeV}$.}
 \end{figure}
\end{center}

Our results are summarized in Figs. \ref{fig:k02}, \ref{fig:k06}, where we show the masses of the first two scalar and the isolines of $\xi$ in the allowed parameter space. For the lightest scalar,
$\xi \lesssim 0.5$ in most of the parameter space, while this is not the case for the next-to-lightest scalar,
for which $\xi \gtrsim 0.3$. The reasons for such a
behavior are rather clear and are summarized in Table \ref{Tab:xi}, where we present the relevant quantities choosing some representative points in parameter space.
\par
The lightest scalar $s_1$ has a reduced coupling to $t\bar{t}$ that suppresses the production cross section via gluon-gluon fusion, and at the same time
it prefers to decay into $b\bar{b}b\bar{b}$ (through the intermediate decay into a pair of lightest pseudoscalars) or into a pair of neutralinos $\chi_1\chi_1$. On
the contrary, $s_2$ has a production cross section much more similar to the Standard Higgs boson (in some cases even higher) that compensates possible depletions in the branching
fraction into vectors due to decays into $A_1 A_1$ or $\chi_1\chi_1$, so that the overall effect is to increase the value of $\xi$ with respect to the one of the lightest scalar.
\section{Conclusions}
With an integrated luminosity of about $2\;{\rm fb}^{-1}$, the LHC is probing ample mass regions not only of the Standard Higgs boson, but also of many interesting
SM extensions. In this paper we focused on $\lambda$SUSY with a scale invariant superpotential, showing the typical
values of $\xi$ (the key quantity probed by the experiments) for the lightest and the next-to-lightest scalars. 
\par
The message is quite clear: the next-to-lightest
scalar is more similar to a standard Higgs boson than the lightest one, and gives in general larger values of $\xi$, although its behavior depends on $k$,
the cubic self coupling of the Singlet.
Since the sensitivity of the LHC in the vector
channels has allowed to already probe values down to $\xi \simeq (0.5\div 0.6)$ \cite{CMS-Atlascomb}, it is clear that the parameter space is becoming more and more constrained.
Our plots seems to indicate that higher values of $k$ are by now disfavored, while some regions in parameter space are still allowed for lower values of $k$, so that
the Peccei-Quinn symmetric limit $k=0$ seems to be still viable. It is plausible that the LHC will be able to probe other regions of parameter space in the next two years, so we will learn more on the
nature of the Electroweak symmetry breaking rather soon.

\section*{Acknowledgments}

We thank Riccardo Barbieri for important suggestions and for reading the manuscript.
This work is supported in part by the European Programme ``Unification in the LHC Era",  contract PITN-GA-2009-237920 (UNILHC). 
E.B. aknowledges support from the Agence Nationale de la Recherche under contract ANR 2010 BLANC 0413 01.

\bibliography{biblio}{}
\bibliographystyle{utphys}

\end{document}